\documentclass[fleqn,10pt,twocolumn]{wlscirep}
\usepackage[utf8]{inputenc}
\usepackage[T1]{fontenc}
\usepackage[range-phrase=--,range-units=single]{siunitx}

\usepackage{amsmath,amssymb}
\usepackage{graphicx}
\usepackage{bm}

\DeclareSIUnit\invcmc{\cm\tothe{-3}}

\DeclareMathOperator{\sinc}{sinc}

\title{Degenerate optical parametric amplification in CMOS silicon}

\author[1,+,*]{David Heydari}
\author[2,+]{Mircea Catuneanu}
\author[1,3,+]{Edwin Ng}
\author[4]{Dodd J. Gray Jr.}
\author[3,4]{Ryan Hamerly}
\author[1]{Jatadhari Mishra}
\author[1,3]{Marc Jankowski}
\author[1]{M.~M.~Fejer}
\author[2]{Kambiz Jamshidi}
\author[1]{Hideo~Mabuchi}
\affil[1]{E. L. Ginzton Laboratory, Stanford University, Stanford, CA, 94305, USA}
\affil[2]{Integrated Photonic Devices Group, Chair of RF and Photonics, Technische Universität Dresden, 01069 Germany}
\affil[3]{NTT Research Inc., Physics and Informatics Laboratories, 940 Stewart Dr., Sunnyvale, CA 94085, USA}
\affil[4]{Research Laboratory of Electronics, MIT, 50 Vassar Street, Cambridge, MA 02139, USA}
\affil[*]{dheydari@stanford.edu}
\affil[+]{These authors contributed equally.}

\begin{abstract}
Silicon is a common material for photonics due to its favorable optical properties in the telecom and mid-wave IR bands, as well as compatibility with a wide range of complementary metal-oxide semiconductor (CMOS) foundry processes.  Crystalline inversion symmetry precludes silicon from natively exhibiting second-order nonlinear optical processes.  In this work, we build on recent work in silicon photonics that break this material symmetry using large bias fields, thereby enabling $\chi^{(2)}$ interactions.  Using this approach, we demonstrate both second-harmonic generation (with a normalized efficiency of \SI{0.20}{\percent\per\watt\per\centi\meter\squared}) and, to our knowledge, the first degenerate $\chi^{(2)}$ optical parametric amplifier (with relative gain of \SI{0.02}{\dB} using \SI{3}{\milli\watt} of pump power on-chip at a pump wavelength of \SI{1196}{\nano\meter}) using silicon-on-insulator waveguides fabricated in a CMOS-compatible commercial foundry.  We expect this technology to enable the integration of novel nonlinear optical devices such as optical parametric amplifiers, oscillators, and frequency converters into large-scale, hybrid photonic-electronic systems by leveraging the extensive ecosystem of CMOS fabrication.
\end{abstract}
\begin{document}

\flushbottom
\maketitle

\thispagestyle{empty}

\section*{Introduction}
Optical nonlinearities are crucial for modern optical systems, useful for sensing\cite{Villares2014}, computation (both quantum\cite{Chuang1995, Yanagimoto2022} and classical\cite{McMahon2016, Li2022}), optical signal processing\cite{Bogaerts2020}, communication\cite{Lu2016}, and attoscience\cite{Lewenstein1994}.
Many of these applications will benefit from the realization of efficient nonlinear interactions in compact photonic integrated circuits.  
Considerable recent work has focused on conventional platforms for integrated photonics, such as silicon and silicon nitride, which are mature and readily available in CMOS foundries\cite{SiewReview}.
However, these material systems possess inversion symmetry, and therefore lack a native second-order susceptibility\cite{RazeghiBook}.
Silicon does possess a rather large third-order nonlinear susceptibility\cite{Leuthold2010, Kuyken2017}, which has been used to demonstrate a number of useful applications including supercontinuum generation\cite{Lafforgue2022}, third-harmonic generation\cite{Corcoran2010}, and optical amplification via stimulated Raman scattering\cite{Claps2003} or non-degenerate four-wave mixing\cite{Foster2006}; however, the power requirements for such demonstrations have been quite high.
For these reasons, tremendous efforts have been devoted to researching materials with native second-order nonlinearities\cite{Xiong2012, Zhang2021, Koos2016, Chang2018}.

\begin{figure*}
\includegraphics[width=\textwidth,height=\textheight,keepaspectratio]{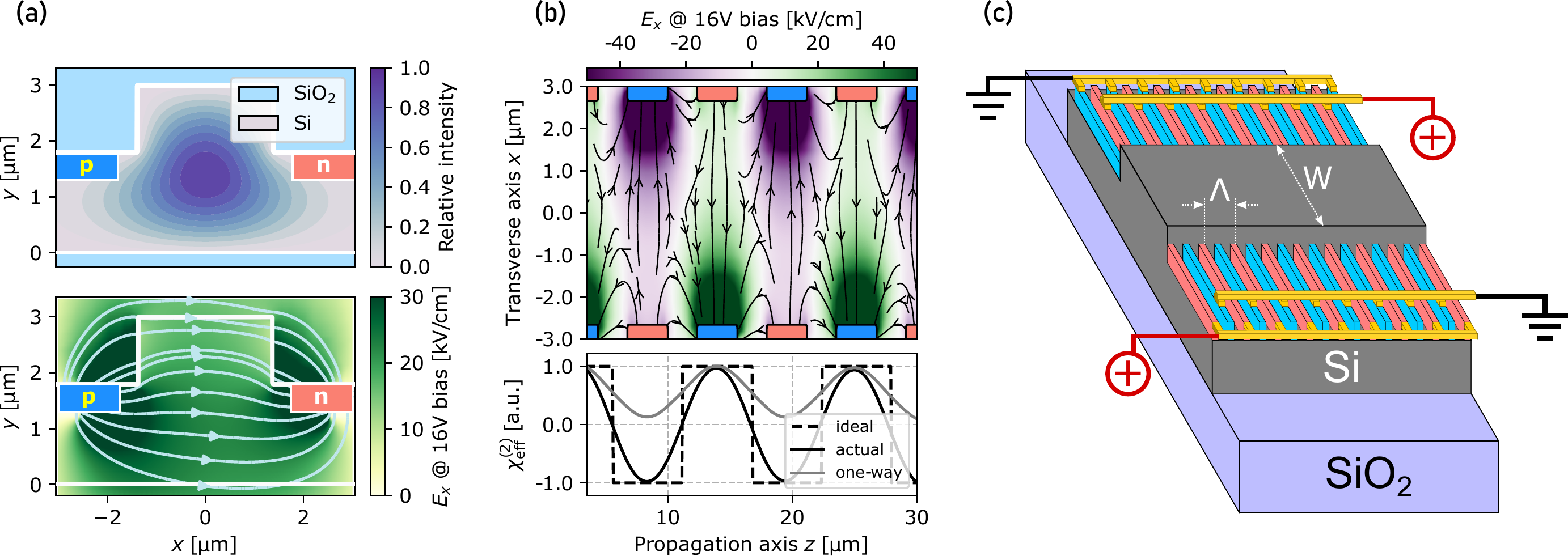}
\caption{Schematics of the device design for DC-field-induced second-order nonlinearity in silicon waveguides. \textbf{(a)} Top: Electric-field profile of TE$_{00}$ mode in waveguide, Bottom: DC-electric field profile along $x$-axis produced by p-i-n diode biased at \SI{16}{\volt} with equipotential lines drawn as arrows; \textbf{(b)} Top: DC-electric field profile along $x$-axis at \SI{16}{\volt} bias as a function of the propagation axis $z$, Bottom: Simulated normalized $\chi_{\rm{eff}}^{(2)}$ profile as function of $z$ for the ideal, two-way poled, and one-way pole cases; \textbf{(c)} Diagram of the silicon rib waveguides with alternating p-i-n diode structures and bias points.}
\label{fig:design}
\end{figure*}

In principle, a second-order nonlinearity may be realized in silicon either by straining the crystalline lattice or by applying an electric field\cite{Jacobsen2006,Chmielak2013,Borghi2015}.
The latter approach, where a large DC bias field converts the native third-order $\chi^{(3)}$ nonlinearity into an effective second-order $\chi^{(2)}$ nonlinearity, is responsible for the well-known phenomenon of electric-field-induced second-harmonic generation (EFISH)\cite{Lee1967}.
Crucially, a sign change of the applied bias field results in a corresponding sign change of the generated $\chi^{(2)}$, which allows for arbitrary nonlinear interactions to be quasi-phasematched\cite{Fejer1992}; the phase-mismatch between the interacting waves can be compensated with a periodic sign change of the $\chi^{(2)}$.
Given the densely integrated systems available from silicon photonics foundries and the diverse functionalities enabled by second-order nonlinear interactions\cite{Jankowski2021}, the realization of quasi-phasematched $\chi^{(2)}$ interactions in silicon promises the best of both worlds.
To date, experimental demonstrations have focused exclusively on second-harmonic generation (SHG), which is commonly used to benchmark the normalized efficiency of a nonlinear waveguide\cite{timurdogan}.  

The same process that produces SHG can also be used for degenerate optical parametric amplification (OPA).  
With a strong second-harmonic pump, the device can be used to provide phase-sensitive amplification to signals at the fundamental harmonic, providing a source of quantum-limited optical gain.  
With sufficient gain, OPA integrated into a low-loss resonator constitutes an optical parametric oscillator, which could in principle be used to realize a source of broadly tunable coherent light in integrated silicon photonics.  
Finally, OPA can also facilitate the generation of quadrature-squeezed vacuum, a foundational resource for photonic quantum information processing schemes such as heralded single-photon sources, cluster-state quantum computation, quantum key distribution, and enhanced metrology beyond the standard quantum limit\cite{Choi1997}.  
This will produce, in effect, a rich, new quantum optics platform in silicon\cite{Rosenfeld2020}.

Thus far, demonstrations of optical amplification in silicon have relied on Raman amplification schemes or third-order Kerr nonlinearities involving complicated four-wave mixing processes\cite{Liu2010, Wang2015, Ooi2017} with high input power requirements.  
Here, for the first time, we demonstrate degenerate, phase-sensitive parametric amplification using second-order nonlinearities in a CMOS platform.  
We first characterized the nonlinearity of these devices using SHG and measured a normalized efficiency of \SI{0.2}{\percent\per\watt\per\centi\meter\squared}.  
Then, we optically pump these waveguides with the second-harmonic and measure phase-sensitive relative power gains as large as \SI{0.02}{\dB} with approximately \SI{3}{\milli\watt} of pump power on-chip at a pump wavelength of \SI{1196}{\nano\meter} and a nonlinear gain section of about \SI{7.5}{\milli\meter}.

\section*{Results}
\subsection*{Chip design}
The waveguides were fabricated at VTT, a commercial foundry specialized in processing \SI{3}{\micro\meter}-thick silicon\cite{Aalto2019}.  
This unique platform, unlike standard \SI{220}{\nano\meter} processes, enables very low loss waveguides (scattering losses $\approx$ \SI{0.1}{\deci\bel\per\centi\meter}), as the mode is largely confined to the waveguide core.
While the larger modal areas reduce the effective nonlinearities achievable in the platform, this disadvantage is offset by the lower intrinsic losses as well as a number of additional practical advantages.
First, the transverse mode profiles are more radially symmetric and better mode-matched to optical fiber, in principle reducing coupling losses suffered at the chip-fiber interface.
Furthermore, compared to more tightly confining waveguides, the waveguide dispersion in these devices are closer to that of bulk silicon, which reduces the index difference between the interacting waves and permits the use of longer quasi-phase matching (QPM) periods.
This results in larger separation between the diodes used to produce the DC electric fields (Fig.~\ref{fig:design}c) and reduces the formation of fringing fields, which in turn mitigates premature longitudinal breakdown while also bringing the longitudinal profile of the induced nonlinearity (i.e., the polarity of the nonlinearity) closer to an idealized square-wave QPM pattern.
Finally, longer QPM periods also reduce the sensitivity of the phase-matching wavelength to fabrication errors\cite{Fejer1992, Jankowski2021}, making our device parameters more robust than those for standard \SI{220}{\nano\meter} processes.

We denote the field amplitudes for the second-harmonic (SH) and fundamental-harmonic (FH) modes by $A_\text{SH}(z)$ and $A_\text{FH}(z)$, respectively, which we assume vary only with the position $z$ along the waveguide.  
These amplitudes correspond to electric fields $A_i(z) \bm{E}_i(x,y) e^{-j \beta_i z}$ ($i = \text{SH},\text{FH}$), where $\bm{E}_i(x,y)$ are the transverse profiles of the modes (see Fig.~\ref{fig:design}a) and $\beta_i = 2\pi n_i/\lambda_i$ are their propagation constants given their wavelengths $\lambda_i$ and modal indices $n_i$.  In this work, we normalize the fields to satisfy
\begin{equation}
    P = \frac{1}{2} \int{\bigl(\bm{E}_i(x,y) \times \bm{H}^{\ast}_i(x,y)\bigr) \cdot \hat{\bm{z}}\, \,\mathrm{d}x\,\mathrm{d}y} = \SI{1}{\watt},
\end{equation}
so that the optical powers in the fields are given by $P_i = |A_i|^2 P$.
In the presence of a second-order optical nonlinearity and linear power losses $\alpha_i$, but neglecting two-photon (TPA) or free-carrier absorption (FCA), the propagation of these field amplitudes obey the coupled-wave equations
\begin{subequations} \label{eqn:cme}
\begin{align}
    \partial_z A_{\text{SH}} &= -j \kappa A_{\text{FH}}^2 e^{j \Delta\beta z} - {\textstyle\frac 1 2} \alpha_{\text{SH}} A_{\text{SH}}, \\
    \partial_z A_{\text{FH}} &= -j \kappa A_{\text{SH}} A_{\text{FH}}^{\ast} e^{-j \Delta\beta z} - {\textstyle\frac 1 2} \alpha_{\text{FH}} A_{\text{FH}},
\end{align}
\end{subequations}
where $\Delta \beta = \beta_{\text{SH}} - 2 \beta_{\text{FH}} - 2\pi/\Lambda$ is the phase mismatch given the poling period $\Lambda$, and $\kappa$ is the nonlinear coupling defined as ${\kappa = \epsilon_0 \omega_0 P^{-1} \int_0^\Lambda \iint \bm{E}_{\text{SH}} \cdot \bm{d}(z) : \bm{E}^{\ast}_{\mathrm{FH}} \bm{E}^{\ast}_{\mathrm{FH}} \,\mathrm{d}x\,\mathrm{d}y\,\mathrm{d}z/\Lambda}$.
In the case where all fields are polarized in the $\hat{\bm{x}}$ direction (i.e., for TE$_{00}$ modes) and thus sample only the $\chi^{(3)}_{1111}$ of silicon\cite{Hon2011}, the second-order nonlinear tensor $2\bm{d} = \bm{\chi}_\text{eff}^{(2)}$ induced by a periodically poled DC electric field $E_\text{DC}(z) \hat{\bm{x}}$ simplifies to\cite{timurdogan}
\begin{equation} \label{eq:chi2-eff}
    \chi^{(2)}_{\mathrm{eff}} = 3\chi^{(3)}_{1111} E_{\mathrm{DC}}.
\end{equation}
This induced $\chi^{(2)}_\text{eff}$ scales linearly with the reverse-bias DC voltage, up until longitudinal breakdown between adjacent diodes.
We note that the reverse bias also aids in sweeping out free carriers generated by TPA, significantly reducing the effective free-carrier lifetime and associated nonlinear loss\cite{dodd}.

In silicon nanophotonics, the DC electric fields can be generated using ion-implanted or diffusion-doped p-i-n diodes, and the placement of these diodes can be precisely controlled using photolithography to achieve the specific poling period $\Lambda$ needed to efficiently phase match the desired wavelengths of operation.  
This poling is conventionally done one-way, i.e., without interchanging the polarity of the diodes\cite{timurdogan}, in order to avoid premature longitudinal breakdown across the shorter poling periods needed to phase match the more dispersive waveguides in standard \SI{220}{\nano\meter}-thick silicon.  
However, in one-way poling, the sign of $E_\text{DC}(z)$ does not change between adjacent diodes, which results in a decreased effective nonlinearity, as can be seen via Eq.~\eqref{eq:chi2-eff}.  
Furthermore, at smaller poling periods, there can be residual fringing fields between adjacent diodes which limit the achievable modulation contrast in $E_\text{DC}(z)$.  
These issues can be mitigated by interchanging the diodes midway between poling periods, which can quadruple the conversion efficiency, since the respective Fourier amplitude in $\kappa$ doubles\cite{Franchi}; see Fig.~\ref{fig:design}(b), bottom, for our two-way poled design compared to a one-way poled scenario.

The waveguides were designed to operate at a nominal FH wavelength of $\lambda_\text{FH} = \SI{2400}{\nano\meter}$ (SH wavelength of $\lambda_\text{SH} =\SI{1200}{\nano\meter}$).  
As can be seen in Fig.~\ref{fig:opa_heatmap}, this operating wavelength was chosen as it avoids excessive losses due to free-carrier absorption of the FH, even for substantial free-carrier lifetimes of \SI{100}{\pico\second}, while simultaneously maximizing the wavelength-dependent conversion efficiency\cite{dodd,Soref1987}.
Using an eigenmode solver (Lumerical MODE), we computed the modal index difference $\Delta n$ between SH and FH, from which we determined the poling period that would allow for efficient quasi-phase matching: $\Lambda = 2 \lambda_\text{FH}/\Delta n$ (Fig.~\ref{fig:design}a).  
We designed for a poling period $\Lambda$ of \SI{11.18}{\micro\meter}, which corresponds to a waveguide geometry that guarantees single-mode operation.
In our design, we used two different finite-element electrostatic Boltzmann-transport equation solvers (Sentaurus TCAD and Lumerical Device) to calculate the full 3D profiles of the DC field (Fig.~\ref{fig:design}b) and hence the effective nonlinearity of the devices; based on these calculations, we predicted a normalized SHG efficiency $\eta_0 = \kappa^2 = \SI{0.24}{\percent\per\watt\per\centi\meter\squared}$.

\subsection*{Second harmonic generation}
To characterize the nonlinear-optical performance of these devices, we first measure their SHG conversion efficiency as a function of input FH wavelength.
From these measurements, we can extract the peak phase-matching wavelength and the normalized SHG efficiency $\eta_0=\kappa^2$.

\begin{figure}[ht!]
\centering
\includegraphics[width=0.95\linewidth]{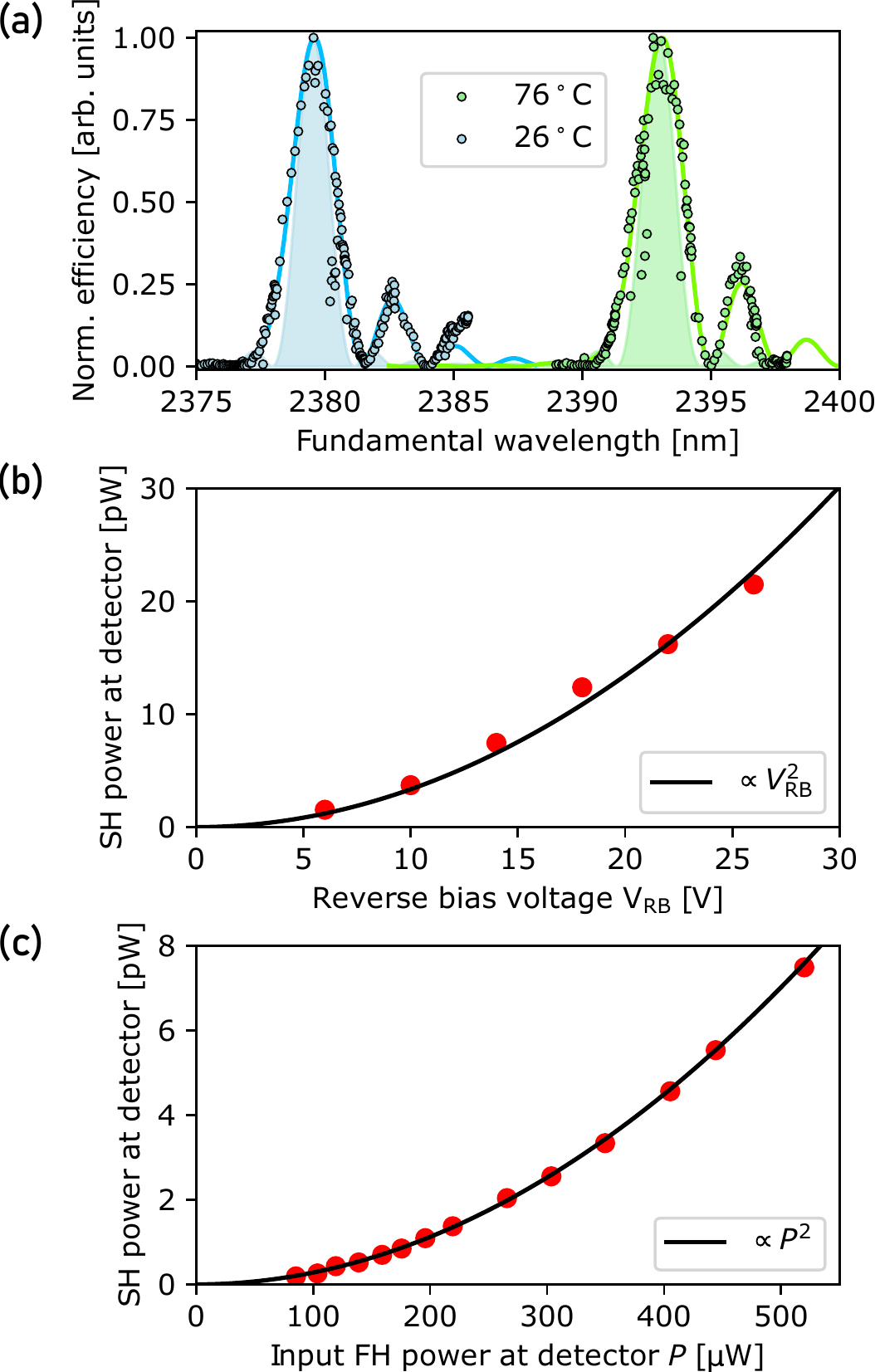}
\caption{Second-harmonic generation via induced second-order optical nonlinearity. \textbf{(a)} SHG efficiency spectrum versus fundamental wavelength taken at two chip temperatures, showing data (dots), theory assuming no fabrication variations (shaded), and theory accounting for geometry variations (solid line); \textbf{(b)} SH power collected at output detector versus reverse-bias voltage (dots) with quadratic curve fit (black solid line); \textbf{(c)} SH power collected at detector versus input FH power at detector (dots) with quadratic curve fit (black solid line).}
\label{fig:results_shg}
\end{figure}

A tunable optical parametric oscillator (Toptica TOPO) generated FH light with wavelengths ranging from 2300 to \SI{2400}{\nano\meter} (Fig.~\ref{fig:results_shg}a), which was coupled into the waveguide through a free-space objective (Newport 5726-C-H).
To activate the nonlinearity, we used a voltage source to supply a reverse bias across the diodes.
Above \SI{31}{\volt} reverse bias, the device began to exhibit characteristics of premature breakdown due to the presence of an undesired shunt via neighboring electrical contacts.  
Therefore, the applied bias voltage was set to \SI{16}{\volt} for the SHG experiments.
We collected the output light from the waveguide using a reflective objective (Thorlabs LMM-15X-P01), which enabled simultaneous collimation of the FH and SH beams.
The collected beams were sent through a pellicle beamsplitter (Thorlabs BP145B1) acting as a pickoff to two detectors: one measured the FH power $P_\text{FH}$ (Thorlabs PDA10D2) through a germanium window (Thorlabs WG91050-C9) to block the SH, while the other measured the SH power $P_\text{SH}$ (Newport Model 2153) through a shortpass filter (Edmunds Optics 84-656) to block the FH.
From this data, we calculated the relative SHG efficiency (transfer function) $P_\text{SH}/P_\text{FH}^2$ as a function of wavelength, which showed the expected $\sinc{}^2$ dependency (Fig.~\ref{fig:results_shg}a).
We also verified that the SHG transfer function was temperature sensitive as expected from the thermo-optic effect, with a tuning rate of \SI{0.26}{\nano\meter/\celsius}, which agrees well with calculations using tabulated data\cite{Li1980}.

\begin{figure*}
\includegraphics[width=\textwidth,keepaspectratio]{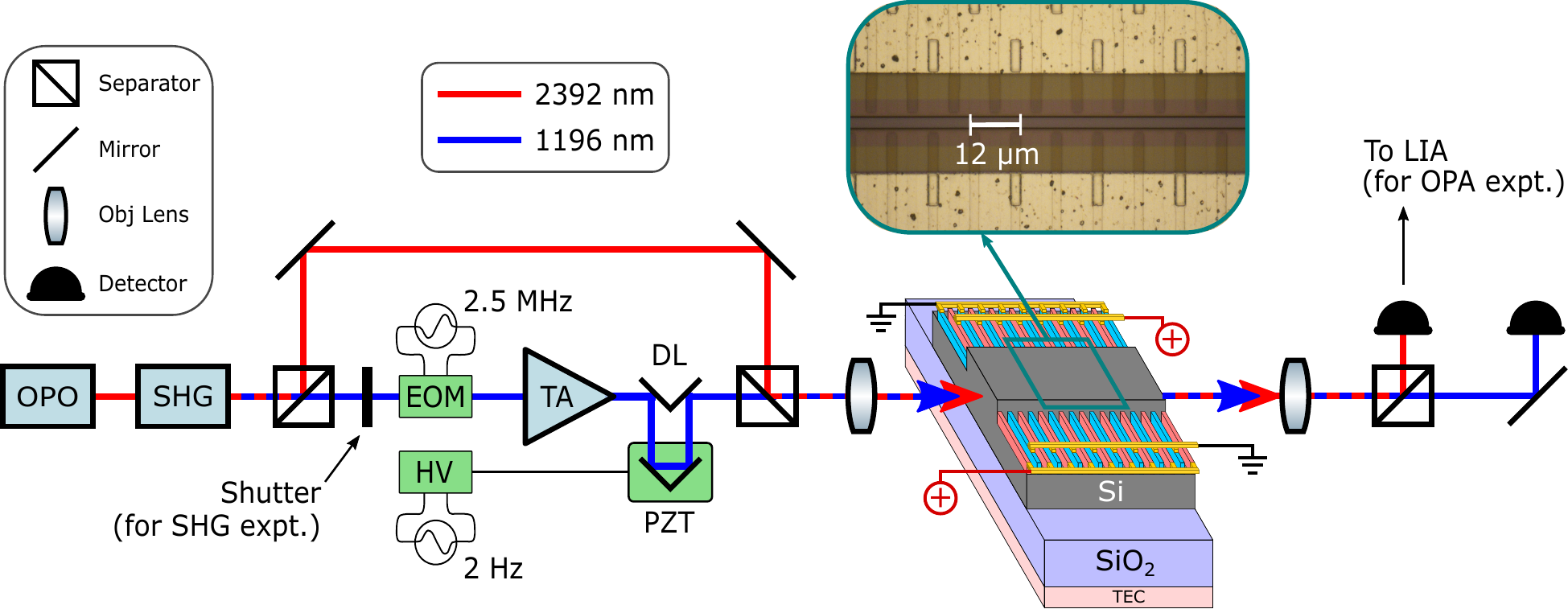}
\caption{Experimental schematic for free-space second-harmonic generation and optical parametric amplification with control electronics for slow-scan/fast-dither lock-in detection.  A shutter is used to separate the two experimental modes (SHG and OPA).  OPO: Tunable optical parametric oscillator; SHG: second-harmonic generation crystal; EOM: electro-optic modulator; PZT: piezo-transducer; HV: high-voltage PZT driver; TA: tapered amplifier; DL: optical delay line; TEC: thermoelectric cooler; LIA: lock-in amplifier.}
\label{fig:schematic}
\end{figure*}

The SHG transfer functions in Fig.~\ref{fig:results_shg}a demonstrated three anomalies: the peak wavelength is not exactly at the design wavelength, the simulated transfer function has a narrower bandwidth than the data, and there appears to be an asymmetric sidelobe in the spectrum.
The first two effects could be explained by fabrication imperfections in any of the geometric properties of the waveguide, such as etch depth, silicon thickness, and width.
Since the geometry of the waveguide influences the dispersion and hence phase-matching conditions, random or systematic variations in these parameters can in theory broaden the transfer function or shift its peak; our simulations indicate that the observed deviations are consistent with the known fabrication variations or non-uniformities in VTT's processes.
The asymmetric sidelobes do not disappear even when changing the chip temperature.
The sidelobes could also be due to phase matching into another mode; however, the latter is less likely, for two reasons: (1) these waveguides were designed to be single-mode\cite{Petermann1976, Soref1991}; (2) different modes have different thermo-optic tuning rates, and our data shows that the sidelobe remains red of the main peak with a consistent position in wavelength when temperature tuning.
They could also be explained by geometric variations---an explanation offered in other nanophotonic nonlinear optics works\cite{Umeki2010, Chauvet2016, Neradovskiy2018, Zhao2020}.
In fact, a quadratic variation in $\partial_z\Delta \beta(z)$, potentially corresponding to a nonlinearly varying geometric parameter such as thickness or width, is able to produce a theory curve with sidelobes consistent with those that we observed.
Putting these effects together, we arrived at a model that qualitatively captures most of the spectral features we obtained in the experimental data (Fig.~\ref{fig:results_shg}a, solid lines).
The resulting fit indicates a reduction in the peak of the SHG transfer function to $\num{0.87}\eta_0$.

Absolute SHG efficiency measurements were performed at \SI{26}{\celsius} and at the wavelength observed to maximize the SHG transfer function.
For this experiment, FH light was coupled into a lensed single-mode fiber mode-matched to our waveguide geometry (WT\&T Technology).
The output SH and FH light were then collected on a lensed multimode fiber (WT\&T Technology) which was fiber-coupled independently to two different optical power meters: one measured the generated SH light (Newport 818-IG) and the other the FH (Thorlabs S405C).  
For the SH power measurements, we used a short-pass filter (Edmunds Optics SP1600) to ensure no FH light was being absorbed by the power meter; for the FH power measurements, there was so little SH power that blocking it out did not affect our readings.
As expected, the SH signal disappeared completely when the reverse bias was turned off, confirming that the transmitted FH did not contribute a background to the detected SH.
After accounting for the collection efficiency at the output, we estimated $P_\text{SH}(L) = \SI{25}{\nano\watt}$ and $P_\text{FH}(L) = \SI{4.7}{\milli\watt}$ at the end of the waveguide.

Using the coupled mode equations \eqref{eqn:cme} and the undepleted pump approximation with loss on both the FH and SH (note that $P_{\text{FH}}(z) \approx P_{\text{FH}}(0) e^{-\alpha_{\text{FH}}z}$), we theoretically expect\cite{timurdogan}
\begin{align}
    &P_{\text{SH}}(L) = \eta_0 P^2_{\text{FH}}(0) L^2 e^{-(\alpha_{\text{FH}} + \frac12\alpha_{\text{SH}})L} \label{eqn:Psh}
\\
    &\quad{}\times \frac{\sin^2\left(\frac12\Delta k L\right) + \sinh^2\left(\frac12(\alpha_{\text{FH}} - \frac12\alpha_{\text{SH}})L\right)}{\left(\frac12\Delta k L\right)^2 + \left(\frac12(\alpha_{\text{FH}} - \frac12\alpha_{\text{SH}})L\right)^2},
    \nonumber
\end{align}
for a nonlinear section of length $L$.
As described in \hyperref[sec:methods]{Methods}, we estimate modal losses of $\alpha_{\text{FH}} = \SI{2.8}{\dB\per\cm}$ and $\alpha_{\text{SH}} =  \SI{4.9}{\dB\per\cm}$.
Hence, we determined that the normalized SHG efficiency of our device is approximately $\eta_0 = \SI{0.2}{\percent\per\watt\per\centi\meter\squared}$, in good experimental agreement with our theoretical design prediction of \SI{0.24}{\percent\per\watt\per\centi\meter\squared}.

Further verification of the SHG results included varying the reverse-bias voltage and the input FH power.
According to Eq.~\eqref{eqn:Psh}, the SH power generated should vary as $P_{\text{FH}}^2$.
Furthermore, since $\kappa=\sqrt{\eta_0}$ depends linearly on the reverse-bias voltage $V_{\rm{RB}}$, the SH power generated should vary as $V_{\rm{RB}}^2$.
Both of these trends were observed (Figs.~\ref{fig:results_shg}b and \ref{fig:results_shg}c).

\subsection*{Optical parametric amplification}
Degenerate optical parametric amplification (OPA) is a phase-sensitive process that occurs given a strong SH pump and an small-signal input FH.
Under an undepleted pump approximation, the coupled mode equations predict that the FH signal power can either increase or decrease exponentially.  
In the absence of losses and when the relative phase between pump and signal is $2\pi n \pm \pi/2$ (for $n$ an integer), the signal output power is $P_\text{FH}(L) = P_\text{FH}(0) e^{\pm 2|\gamma|L}$, where $|\gamma| = \sqrt{\eta_0 P_\mathrm{SH}(0)}$.
In other words, the amount of amplification of the signal in the waveguide is modulated by the phase of the pump relative to the signal.
After including losses on the FH and SH, a general expression for the relative gain $g = P_\text{FH}(L)/(P_\text{FH}(0)e^{-\alpha_\text{FH}L})$ (at phase matching) is
\begin{equation}
    g(\phi) = \cosh^2{\Tilde{\gamma} L} + \sinh^2{\Tilde{\gamma} L} + 2 \sinh{\Tilde{\gamma} L} \cosh{\Tilde{\gamma} L} \sin{\phi},
\end{equation}
where $\phi$ is the relative phase difference between the pump and signal fields\cite{Jankowski2021} and $\Tilde{\gamma} = |\gamma| e^{-\frac12\alpha_{\text{SH}}L}$.  In the limit of small amplification where $\Tilde\gamma L \ll 1$,
\begin{equation} \label{eq:opa-small-gain}
    g(\phi) = 1 + 2\Tilde{\gamma}L \sin{\phi}.
\end{equation}
Therefore, in this limit, the gain scales as the square root of both the SH pump power and the normalized efficiency.

In order to generate phase-coherent light at both FH and SH for this experiment, we frequency-doubled the TOPO idler output using a bulk SHG setup to produce seed pump light at the FH, which was then amplified by a tapered amplifier (TA, Toptica BoosTA PRO 1200), as shown in Fig.~\ref{fig:schematic}.
The nominal operating wavelength for these experiments was around \SI{1196}{\nano\meter} SH (\SI{2392}{\nano\meter} FH) as the \SI{3}{\deci\bel} operating bandwidth of the TA was limited to 1190--\SI{1210}{\nano\meter}.

To measure gain and observe its phase sensitivity, we varied the relative pump phase as a function of time and measured the response of the signal.
The experiment utilized a slow phase scan combined with a fast phase dither.
The slow phase scan produced multiple full swings through amplification and de-amplification, which helped reduce the impact of fluctuations in the optical path length due to air currents, and it consisted of a free-space delay line which scanned the pump phase using a retroreflector displaced by a piezoelectric transducer (PZT).

The PZT was driven by a high-voltage amplifier (Thorlabs MDT693B) which provided a triangle waveform (sourced from a Tektronix AFG3252C) at frequency $f_\text{r} = \SI{2}{\hertz}$; beyond this frequency, the PZT/retroreflector setup was unable to drive the delay line linearly.
We denote this linear scan of the phase by the function $\phi_\text{ramp}(t)$.
In addition, a fast phase dither was also employed to facilitate lock-in detection of the relatively small gain, and it was produced by a fiber phase modulator (iX-Blue NIR-MPX-LN-10), which imparted a small-signal phase modulation to the fiber-coupled seed light sent to the TA.
We used a sinusoidal modulation with frequency $f_\text{d}=\SI{2.5}{\mega\hertz}$ and a modulation depth $\delta$.
In total, our scheme used a time-varying pump phase $\phi(t) = \phi_\text{ramp}(t) + \delta \cos(2\pi f_\text{d} t)$.

The TA produced about \SI{0.6}{\watt} of phase-modulated SH pump light, which was combined with the FH signal beam at a dichroic mirror (Layertec 103080), and both beams were co-aligned into the waveguide.
The experiment was optimized for coupling of the pump at the expense of that of the signal.
Based on the SH power measured at the output, we estimated an input coupling efficiency of about \SI{0.5}{\percent} for the pump.

During the OPA experiments, the device was observed to generate a photocurrent of approximately \SI{1.5}{\milli\ampere}, which suggests the device absorbed at least \SI{1.5}{\milli\watt} of SH pump light (assuming a quantum efficiency [incident photon-to-converted-electron ratio] of 1).
This could be due to a portion of the input pump light being absorbed by the heavily doped regions that flank the intrinsic core, which are known to possess modified electronic properties such as a shrinkage of the band gap\cite{Schmid1981, Hava1993}.
We observed that the photocurrent is wavelength dependent, as it was virtually absent at the FH signal wavelength.
It was also observed to vary linearly with pump power, ruling out a nonlinear carrier-generation process such as TPA-induced FCA.
It is therefore likely that absorption losses on the FH are dominated by FCA, which does not result in a photoresponse, while the absorption losses on the SH include extrinsic impurity level-to-band absorption, which can result in a photoresponse\cite{Schroder1978}. We therefore augmented our estimates for the SH loss accordingly (see \hyperref[sec:methods]{Methods}).
Given the steep exponential dropoff of the transverse SH mode profile, placing the dopants slightly further away from the core can significantly reduce this loss in future device designs, without significant reduction in the DC electric field and hence nonlinearity.

After the waveguide, the output beams were collected using a reflective objective (Thorlabs LMM-15X-P01), which enabled collimation of both FH and SH beams and was especially useful for input alignment because it enabled simultaneous imaging of both input beams on a camera (FLIR A6700).
For measurements, we directed the collected light through a pellicle beamsplitter (Thorlabs BP145B1) acting as a pickoff, and the majority of FH power was measured by a detector (Thorlabs PDA10D2) located behind a germanium window (Thorlabs WG91050-C9) to block the SH beam.
The output signal from this detector was then fed into a lock-in amplifier (SRS SR865A) referenced to the dither signal at $f_\text{d}$.  Under the gain predicted by Eq.~\eqref{eq:opa-small-gain}, the lock-in output should follow
\begin{equation}
    \sqrt{2} V_{\text{LIA}} = \overline{V}_\text{FH} (g_\text{max}-1) \delta \cos\phi_\text{ramp}(t),
\end{equation}
where $g_\text{max} = g(\pi/2)$ and $\overline{V}_\text{FH}$ is the nominal (DC) voltage at the signal detector in the absence of pump light and hence OPA.  To obtain clean signals, we set the built-in noise filter (standard high-order RC) on the lock-in to \SI{24}{\dB/oct}.

\begin{figure}[h!t]
\centering
\includegraphics[width=0.95\linewidth]{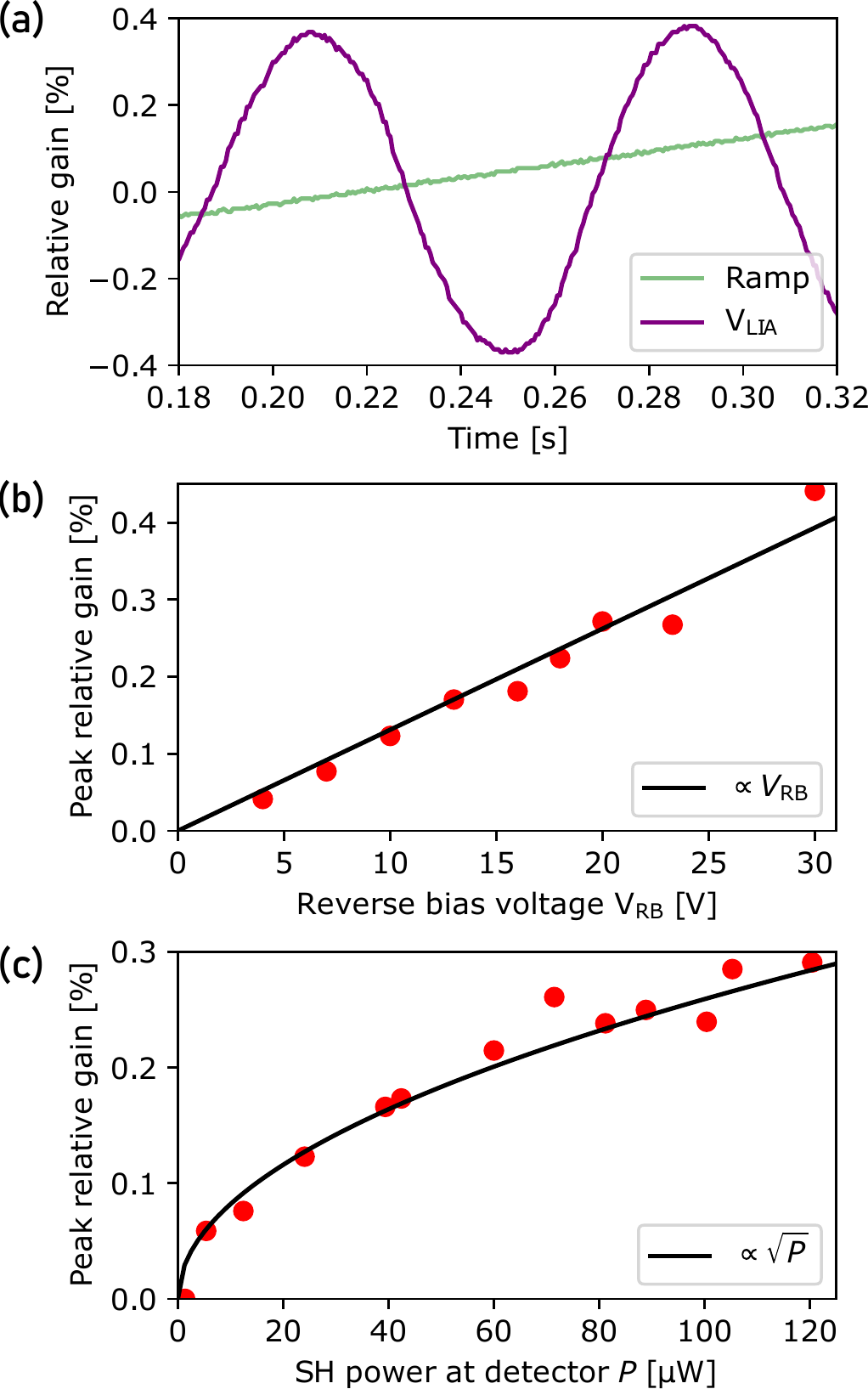}
\caption{Degenerate optical parametric amplification via induced second-order optical nonlinearity. \textbf{(a)} Swing of the OPA gain $g(\phi)-1$ as function of the input pump phase which is slowly scanned in time; \textbf{(b)} Peak relative OPA gain (computed by averaging the peaks observed over one period of the slow phase scan) as a function of reverse bias voltage (dots) with linear fit (solid black line); \textbf{(c)} Peak relative OPA gain as function of SH pump power, measured at output detector, with square-root fit (solid black line).}
\label{fig:results_opa}
\end{figure}

In Fig.~\ref{fig:results_opa}a, we show the lock-in output in terms of $\sqrt{2} V_\text{LIA}/\overline{V}_\text{FH}\delta$, as we scanned the pump phase with a reverse-bias voltage at the upper limit of \SI{31}{\volt} to maximize gain. The result showed clear signs of phase sensitivity, as the gain swept between both positive and negative values as a function of the nominal pump phase, indicating both amplification and deamplification of the signal, respectively.
We note there is some drift in the DC offset level of the LIA output on the order of \SI{10}{\percent} of the peaks, with a correlation time on the order of \SI{100}{\milli\second}; this can lead to slight asymmetries between the maximum amplification and deamplification from trace to trace.
This phase-sensitive signal disappeared when the reverse bias was turned off and is thus directly caused by the induced optical nonlinearity.
To further verify that our observations are consistent with OPA gain, the reverse-bias voltage and SH input power were varied.  
According to our gain model, in the absence of nonlinear losses, the gain varies linearly with the reverse-bias voltage and with the square root of the SH input power, which were both observed as shown in Figs.~\ref{fig:results_opa}b and \ref{fig:results_opa}c, respectively.

Using Eq.~\eqref{eq:opa-small-gain}, the estimated losses for FH and SH (see above discussion and \hyperref[sec:methods]{Methods}), and the $\eta_0$ measured in our SHG experiments, we can infer that approximately \SI{3.2}{\milli\watt} of SH power at the beginning of the waveguide would be needed to produce the observed gain of, e.g., \SI{0.2}{\percent} at \SI{16}{V} (see Fig.~\ref{fig:results_opa}b). 
This inferred pump power is consistent with reasonable estimates for the coupling efficiency of SH light into the waveguide as well as measurements made of the collected SH power after the waveguide, under reasonable estimates of the collection efficiency.

Finally, because of the large pump powers involved in OPA, TPA-induced FCA can in principle also cause nonlinear optical loss, particularly of the amplified signal, when the TPA-induced carrier concentrations exceed \SI[retain-unity-mantissa = false, scientific-notation = true]{e18}{\per\centi\meter\cubed}. 
TPA-induced FCA is a function of the carrier lifetime and the optical intensity in the waveguide\cite{Gajda2011, Dimitropoulos2005}.
Still, these carriers can be swept out by the applied DC field, which reduces this loss.
We estimate the free carrier lifetime for these waveguides at \SI{16}{\volt} reverse bias to be approximately \SI{25}{\pico\second} \cite{Turner-Foster2010a, Gajda2011, Hamerly2018, dodd}, whereas a free-carrier lifetime on the order of \SI{}{\micro\second} would be required to generate enough free carriers to produce noticeable TPA-induced FCA loss. Thus, it is unlikely that OPA is affected by nonlinear losses in our operating regime.

\section*{Discussion}
\begin{figure}[ht]
    \centering
    \includegraphics[width=\linewidth]{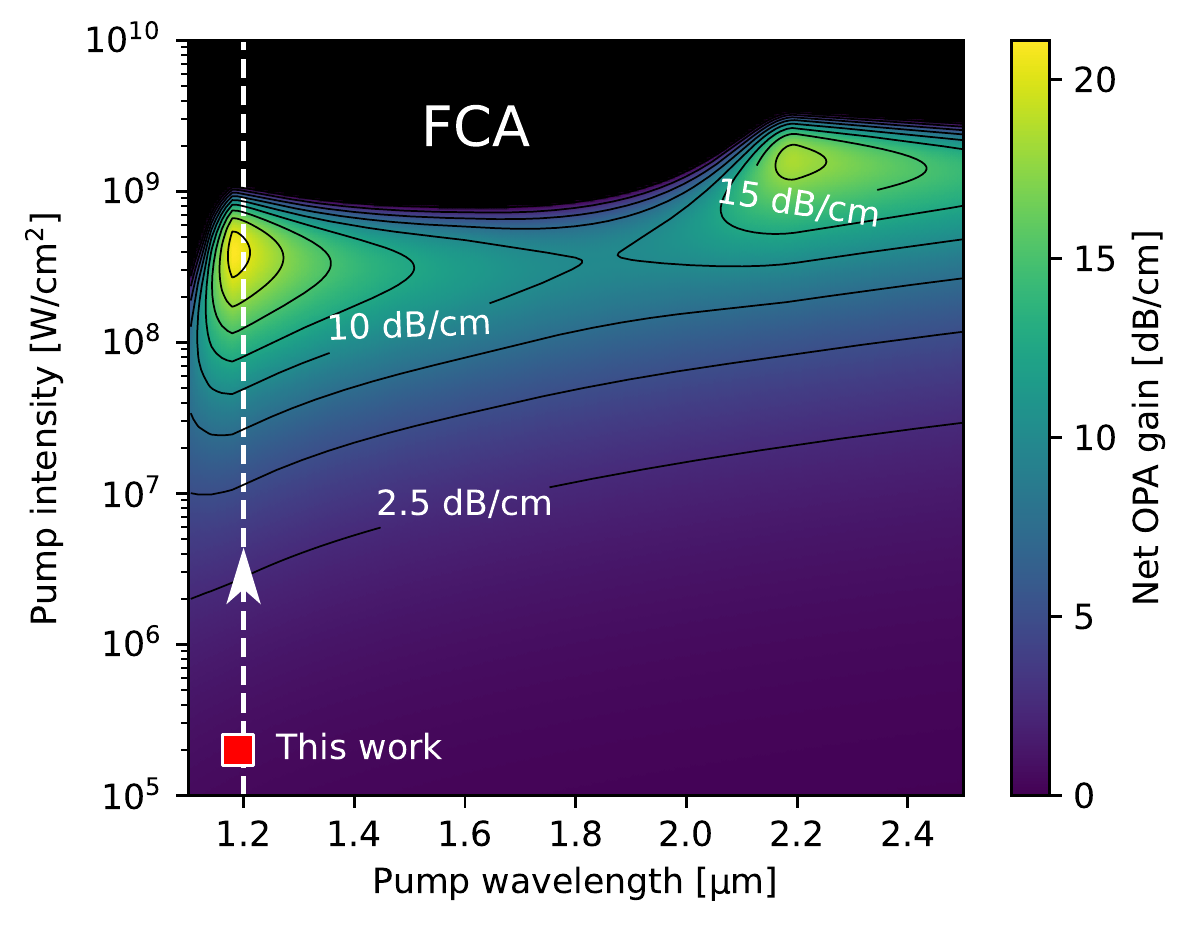}
    \caption{Material-limited net OPA gain (assuming two-way poling and bias field near breakdown of \SI{250}{\kilo\volt\per\centi\meter}) plotted as a function of wavelength and pump intensity for a free carrier lifetime of \SI{100}{\pico\second}. Loss calculations include cross-two-photon absorption, and free-carrier absorption induced from two-, three-, four-, and five-photon absorption.  Dashed line shows optimum wavelength choice for experiments based on gain landscape and potential future trajectory for experiments.  The region in black is associated with negative net OPA gain due to excessive losses from free-carrier absorption (FCA).}
    \label{fig:opa_heatmap}
\end{figure}

In Fig.~\ref{fig:opa_heatmap}, we plot the material-limited net gain as a function of wavelength and pump intensity in the presence of various nonlinear loss mechanisms, including cross-2-photon\cite{Hon2011} absorption and \textit{n}-photon-absorption-induced free carrier absorption\cite{Gai2013}.

The calculation is material-limited in that it is referenced to the OPA gain expected at breakdown and it does not consider waveguide geometry; hence the transverse field profile of the interacting waves is constant.
This provides us with the intrinsic material limits to second-order nonlinear optics with SOI technology, and a roadmap to the next experimental milestone after amplification---oscillation.  
The gain landscape shows two local maxima for pump photon energies near the band edge of silicon ($\lambda_{E_g} \approx \SI{1.15}{\micro\meter}$) and around \SI{2.2}{\micro\meter}.  
We chose to operate in the shorter wavelength regime due to (1) less availability of good detectors at long wavelengths and (2) increased oxide absorption and substrate leakage through the buried oxide layer of the SOI stack.
Fig.~\ref{fig:opa_heatmap} reveals that we could have obtained substantially more gain had we been able to couple more pump light into the waveguide and apply a larger reverse-bias voltage to the diodes.

In conclusion, we have demonstrated the first degenerate $\chi^{(2)}$-based optical parametric amplifier in CMOS-compatible silicon, creating a new testbed for various applications of nonlinear optics using a ubiquitous material that has established itself on an industrial level.  
This was achieved using components readily available in the silicon nanophotonics toolbox such as waveguides and p-n diodes.
Future work could explore using pulsed sources rather than CW ones, as the former can produce considerably higher peak power on the scale of \SI{}{\kilo\watt}, assuming the waveguide is properly dispersion engineered.
We expect this technology to bring $\chi^{(2)}$ nonlinear optics, via a versatile and cost-effective platform, to a variety of fields, including classical and quantum information processing.

\section*{Materials and methods}
\label{sec:methods}
We designed rib waveguides with a nominal width of \SI{2.75}{\micro\meter}, an etch depth of \SI{1.2}{\micro\meter}, and length $L$ of \SI{9.3}{\milli\meter} (poled length of \SI{7.535}{\milli\meter}).  
These were fabricated in a commercial CMOS foundry (VTT) specialized in processing \SI{3}{\micro\meter}-thick silicon.  
Ion implantation was used to dope p- and n-type regions that flanked the intrinsic core \SI{1.4}{\micro\meter} on each side, with each dopant section being \SI{3}{\micro\meter} wide (see Fig.~\ref{fig:design}b for a depiction).

Simulations of the modal loss used electrorefraction and electroabsorption data based on the dopant concentrations used in this platform\cite{Nedeljkovic2011} ($\sim$\SI{1e20}{\per\centi\meter\cubed}) to estimate free-carrier absorption loss, which comes out to $\alpha_\text{FH} = \SI{2.8}{\deci\bel\per\centi\meter}$ for the FH and \SI{1.3}{\deci\bel\per\centi\meter} for the SH.
Based on the observed photocurrent generated by the SH in the OPA measurements, we also inferred additional extrinsic (impurity-doped) loss\cite{Schroder1978} of \SI{3.6}{\deci\bel\per\centi\meter}; thus, we estimated $\alpha_\text{SH} = \SI{4.9}{\deci\bel\per\centi\meter}$.

Two characterization setups were used for the $\chi^{(2)}$ experiments: second-harmonic generation (SHG) and degenerate optical parametric amplification (OPA).
The SHG experiments used a custom lensed fiber mode-matched to the waveguide at the input and a lensed multi-mode fiber at the output (WT\&T Technology); for SHG transfer-function experiments we replaced the multi-mode fiber at the output with a reflective objective (Thorlabs LMM-15X-P01). For OPA experiments, we used a plano-convex aspheric objective at the input (10X, 15.3 mm EFL, 1050--\SI{1600}{\nano\meter} anti-reflection coated, New Focus) and a reflective objective (Thorlabs LMM-15X-P01) at the output.  
On the chip, an optimized taper design for edge coupling ensures low insertion loss as low as \SI{1.5}{\dB} per facet while also preventing mode hybridization.

The waveguides were packaged into custom DIP sockets with fanouts for biasing both the Peltier elements and p-i-n diodes on-chip.  A constant DC electric field was applied to the chip from a voltage supply (Agilent E3630A or Xantrex XHR-300 for voltages exceeding \SI{20}{\volt}).

\bibliography{biblio}

\section*{Author contributions statement}
D.H.\ and M.C.\ designed the chip and built the characterization setup with assistance from E.N. D.H., E.N., and M.C.\ performed the measurements with assistance from J.M. M.J.\ assisted with characterization and data analysis. R.H.\ and D.J.G.\ conceived of and advised the project. H.M., K.J., and M.M.F.\ supervised and secured funding for the project. All authors contributed to writing the manuscript.

\section*{Acknowledgements}
The authors wish to thank NTT Research for their financial and technical support. This work was also supported by the National Science Foundation (NSF) under grants PHY-2011363 and CCF-1918549, and the German Research Foundation (DFG) in the context of the ``Silicon-on-insulator based integrated optical frequency combs for microwave, THz and optical applications'' project (JA 2401/11-2).

\section*{Conflicts of interest}
The authors declare that they have no conflicts of interest.

\end{document}